# Measurement of the $\Omega_c^0$ Lifetime

WA89 Collaboration

hep-ex/9507004   11 Jul 95

# MEASUREMENT OF THE $\Omega_c^0$ LIFETIME

## The WA89 Collaboration

M.I. Adamovich[8], E. Albertson[5,a], Yu.A. Alexandrov[8], D. Barberis[3], M. Beck[5], C. Bérat[4], W. Beusch[2], M. Boss[6,b], S. Brons[5], W. Brückner[5], M. Buénerd[4], C. Büscher[5], F. Charignon[4], J. Chauvin[4], E. Chudakov[7], F. Dropmann[5], J. Engelfried[6,c], F. Faller[6,d], A. Fournier[4], S.G. Gerassimov[5,8], M. Godbersen[5], P. Grafström[2], Th. Haller[5], M. Heidrich[5], R.B. Hurst[3], K. Königsmann[5,e], I. Konorov[5,8], K. Martens[6], Ph. Martin[4], S. Masciocchi[5] R. Michaels[5,f], U. Müller[7,g], C. Newsom[h], S. Paul[5], B. Povh[5], Z. Ren[5], M. Rey-Campagnolle[4], G. Rosner[7], L. Rossi[3], H. Rudolph[7,i], L. Schmitt[5], H.-W. Siebert[6], A. Simon[5,j], V.J. Smith[5,1], O. Thilmann[6], A. Trombini[5,k], E. Vesin[4], B. Volkemer[7], K. Vorwalter[5], Th. Walcher[7], G. Wälder[6,l], R. Werding[5], E. Wittmann[5] M.V. Zavertyaev[5,8].

[1] *University of Bristol, Bristol, United Kingdom.*
[2] *CERN, CH-1211 Genève 23, Switzerland.*
[3] *University of Genova/INFN, I-16146 Genova, Italy.*
[4] *ISN Université Joseph Fourier, IN2P3, F-38026 Grenoble, France.*
[5] *Max-Planck-Institut für Kernphysik#, D-69117 Heidelberg, Germany.*
[6] *Physikalisches Institut der Universität#, D-69120 Heidelberg, Germany.*
[7] *Institut für Kernphysik der Universität#, D-55099 Mainz, Germany.*
[8] *Moscow Lebedev Physics Institute, RU-117924, Moscow, Russia.*

## Abstract

We present the measurement of the lifetime of the $\Omega_c^0$ we have performed using three independent data samples from two different decay modes. Using a $\Sigma^-$ beam of 340 GeV/c we have obtained clean signals for the $\Omega_c^0$ decaying into $\Xi^- K^- \pi^+ \pi^+$ and $\Omega^- \pi^+ \pi^- \pi^+$, avoiding topological cuts normally used in charm analysis. The short but measurable lifetime of the $\Omega_c^0$ is demonstrated by a clear enhancement of the signals at short but finite decay lengths. Using a continuous maximum likelihood method we determined the lifetime to be $\tau_{\Omega_c^0} = 55^{+13}_{-11}(stat)^{+18}_{-23}(syst)$ fs. This makes the $\Omega_c^0$ the shortest living weakly decaying particle observed so far. The short value of the lifetime confirms the predicted pattern of the charmed baryon lifetimes and demonstrates that the strong interaction plays a vital role in the lifetimes of charmed hadrons.

\#) supported by the Bundesministerium für Forschung und Technologie, Germany, under contract numbers 05 5HD15I, 06 HD524I and 06 MZ5265

[a] *Now at ISN Université Joseph Fourier, IN2P3, F-38026 Grenoble, France.*

[b] *Now at SAP AG, D 69190 Walldorf, Germany*

[c] *Now at FNAL, PO Box 500 Batavia, IL 60510, USA.*

[d] *Now at Fraunhofer Inst. für Solar Energiesysteme, D-79100 Freiburg, Germany.*

[e] *Now at Fakultät für Physik, Univ. Freiburg, Germany*

[f] *Now at CEBAF, 12000 Jefferson Ave., Newport News, VA 23606, USA.*

[g] *Now at CERN, Div. PPE, CH-1211 Genève 23, Switzerland.*

[h] *University of Iowa, Iowa City, IA 52242, USA.*

[i] *Now at LBL, MS 50D, Berkeley, CA 94720, USA.*

[j] *Now at Physics Dept., Univ. of New Mexico, Las Cruces, New Mexico, USA*

[k] *Now at LAL, Orsay, France*

[l] *Now at Charmilles Technologies, 1217 Meyrin, Switzerland*

# 1 Introduction

The experimentally observed lifetimes of charmed baryons differ by as much as a factor of three between $\tau(\Xi_c^+) = 350^{+70}_{-40}$ fs [1]-[5] and $\tau(\Xi_c^0) = 98^{+23}_{-15}$ fs [6, 7, 5]. This large variance is understood to be the consequence of the strong interaction between the quarks involved and of W-exchange between the charm quark and a d-quark [8, 9].

If both decay modes are possible, the direct decay of the c-quark by emission of a $W^+$ ($c \to su\overline{d}$) and W-exchange with a d-quark of the baryon ($cd \to su$), the lifetime is short, as is the case for the $\Xi_c^0$. If only direct decay is possible, the lifetime is long, as is the case for $\Xi_c^+$. Further differences among the lifetimes follow from the details of the baryon wavefunctions.

The $\Omega_c^0$ decay is of particular interest. It proceeds dominantly via the direct decay. In the final state, however, three identical quarks are present ($\Omega_c^0 \to sssu\overline{d}$). If the two strange quarks of the $\Omega_c^0$ are only spectators, the $\Omega_c^0$ lifetime will be comparable to the $\Xi_c^+$ lifetime. If, however, the three strange quarks overlap in phase space in the final state, the decay will be accelerated because there are three ways to reach the same final state, and the $\Omega_c^0$ lifetime will be very short. In addition the Cabibbo suppressed W-exchange with an s-quark ($cs \to su$) may contribute [10].

The $\Omega_c^0$ has been observed so far in three different experiments: in the hyperon beam experiment WA62 [11], in the $e^+e^-$ experiment ARGUS [12] and the photon beam experiment E687 [13, 14]. A preliminary estimate of its lifetime has been reported by the collaboration E687 at the HQ94 conference, $\tau$ ($\Omega_c^0$)= 50-100 fs [15], superseded by a result presented at LISHEP 95, $\tau$ ($\Omega_c^0$)= $86^{+27}_{-20}$ fs [16]. In the following, we present a measurement of the $\Omega_c^0$ lifetime from the hyperon beam experiment WA89 at CERN, using data recorded in 1993. Preliminary results were already presented at Moriond 95 [17].

# 2 Beam and apparatus

The hyperon beam is derived from a 450 GeV/c proton beam with an intensity of $4 \cdot 10^{10}$p/spill (2.5 seconds every 14.5 seconds) which interacts in a 40 cm long beryllium target. Negative particles are extracted in the forward direction by a magnetic beam channel of 12 m length defined by a tungsten collimator. Downstream of the collimator, beam particles are detected in a set of 9 silicon microstrip detectors of 50 $\mu$m pitch and two 3 mm thick scintillators used in the trigger. The beam has an intensity of $6 \cdot 10^5$ particles/spill and is composed of $\pi^-$ and $\Sigma^-$ with a ratio of about 2.3:1 at the experiment target. The mean momentum of the $\Sigma^-$ is 340 GeV/c. An 80 cm long transition radiation detector (TRD) serves as a fast tagger on $\Sigma^-$ [18]. The signal of the TRD is used in the 2nd level trigger to reject about 95% of all high energy beam $\pi^-$, keeping 85% of the $\Sigma^-$.

The apparatus is designed as a forward spectrometer and schematically shown in Fig 1. It consists of a target area, a decay zone and a magnetic spectrometer followed by detectors for particle identification and calorimetry.

The beam impinges on a longitudinally segmented target composed of one copper slab (4 mm thick) and 3 carbon plates, made of pressed diamond powder ($\rho \approx 3.3$ g/cm$^3$) with a thickness of 2 mm each. The targets are placed in a closed box flushed with helium and are spaced by 2 cm. A vertex detector made from 29 planes of silicon detectors is placed downstream of the target box. The first 24 planes have a readout pitch of 25 $\mu$m and active areas of 5x5 cm$^2$ and 6.3x6.3 cm$^2$ [19]. The last 5 planes have an active area of 5x5 cm$^2$ and a readout pitch of 50 $\mu$m. Ten detectors measure the horizontal x-coordinate, 10 the vertical y one. Six detectors measure the coordinate (y+45$^0$) and 3 the coordinate (y−45$^0$). Two 3 mm thick scintillators mounted behind the 12th silicon plane are used to trigger on interactions, thus the first part of the vertex detector also serves as silicon targets of 3.6 mm total thickness.

The decay zone downstream of the target area has a length of about 10 m. It contains drift chambers (40 planes) which are interleaved with 1 mm pitch MWPCs with an active area of about 12x12 cm$^2$ including the beam region (20 planes). Their purpose is to detect track segments from hyperon and $K^0$ decays before they enter the magnetic spectrometer. MWPCs with 1 mm pitch (12 planes) placed 2 m downstream of the target are used to link tracks reconstructed separately in the two parts of the detector. The particle momenta and charges are determined in the Omega spectrometer, a large superconducting magnet ($\int B \cdot dl \approx 7.2$ Tm) equipped with MWPCs (45 planes) measuring three different co-ordinates (horizontal and $\pm 10.15^0$). Two sets of large size drift chambers (12 planes) are used to measure the tracks of particles leaving the spectrometer. They are mounted back to back with MWPCs (MY1 and MY2) used for pattern recognition and triggering.

A large Ring Imaging Cherenkov (RICH) detector serves for the identification of particles with momenta above 12 GeV/c [20]. It provides $\pi$/K separation up to 100 GeV/c. Two hodoscopes, H1 and H2, are mounted on either side of the RICH and serve as trigger detectors. Downstream of H2 we use a lead glass calorimeter with an active area of 3 m$^2$ [21] followed by a hadron calorimeter made from lead/scintillating fibres constructed in the spaghetti calorimeter technique [22].

The main trigger consists of three levels. At the first level, a beam trigger and a signal from the two interaction counters corresponding to two or more particles are required. At the second level, signals from $\geq 3$ charged particles are required in H1 and MY1 and hit correlations between H1, H2 and MY1 are used to enrich the sample with events having at least two charged particles above 30 GeV/c. $\Sigma^-$ interactions are selected at the third level using the TRD (see above). This rather loose interaction trigger and the geometrical acceptance of the spectrometer limit the acceptance of the experiment to charmed hadrons produced with $x_f \geq 0.15 - 0.2$, depending on the final state.

Track reconstruction is performed separately in each detector area. Track segments are then linked towards the upstream direction, from the spectrometer to the decay zone and then to the vertex detectors. Track segments in the vertex detector linked to track segments in the spectrometer are refitted taking into account multiple Coulomb scattering. Beam tracks are reconstructed in the beam detectors and are required to match the geometry of the magnetic beam channel to reject particles from hyperon decays or from interactions in the walls of the collimator.

# 3   Selection and Reconstruction of $\Omega_c^0$ Decays

The data sample comprises $1.5 \cdot 10^8$ interactions on tape which correspond to about $2 \cdot 10^{10}$ incoming $\Sigma^-$ and $3.8 \cdot 10^8$ interactions. The decay channels used in this analysis are

$$\Omega_c^0 \rightarrow \Omega^- \pi^+ \pi^- \pi^+ \ (1)$$
$$\Omega_c^0 \rightarrow \Xi^- K^- \pi^+ \pi^+ \ (2)$$

The selection of candidates for charm baryon decays is based on particle identification, reconstruction of the interaction point and isolation of the secondary vertex.

$\Lambda \rightarrow p\pi^-$ are identified through their decay geometry and kinematics. $\Omega^- \rightarrow \Lambda K^-$ and $\Xi^- \rightarrow \Lambda\pi^-$ decays are reconstructed from negative particles forming a vertex with a $\Lambda$ and giving an invariant mass near the nominal mass of $\Omega^-$ or $\Xi^-$. Only combinations are kept in which the momentum sum points to a track in the vertex detector (assumed to be the $\Omega^-$ or $\Xi^-$) which has no matching track in downstream detectors. In order to reduce background in the $\Omega^-$ sample we exclude track combinations with a $\Lambda\pi^-$ mass consistent with a $\Xi^-$. For positively RICH-identified $K^-$ with momentum greater 25 GeV/c the window around the $\Xi^-$ mass is $\pm 5$ MeV/c$^2$, whereas below that value the window is $\pm 10$ MeV/c$^2$ and we require that in addition the $\pi^-$ hypothesis is rejected by the RICH if the $K^-$ candidate falls into its geometrical acceptance. The invariant mass plots for the hyperons used in this analysis are shown in Fig. 2.

The reconstruction of the charm decay vertex follows a 'candidate-driven' approach. First we select track combinations for which the particle types or at least their charges correspond to charm decays. A common vertex then is reconstructed for these tracks ('secondary vertex'). From the remaining tracks we determine the position of the 'main vertex', iteratively omitting tracks from the combination until the quality criterium ($\chi^2/df \leq 5$) is met [23]. At least three tracks are required for a main vertex.

In the following we describe the selection of the charm candidates in more detail. The criteria used are generally the same for all samples allowing small variations of their values depending on the final state. Reconstructed secondary vertices are retained if they fulfill requirements on the $\chi^2$ of the vertex ($\chi^2/df \leq 5$-6). We also require that the calculated position error for the main vertex is below 300-500 $\mu m$ and below 1 mm for the charm decay vertex. Tracks with momenta below $1-2$ GeV/c are not considered for secondary vertices.

In order to obtain clean mass distributions, charm samples are generally selected using also other criteria, not used for the final lifetime analysis. These are based on the topologies of these two vertices (e.g. vertex isolation and individual track impacts) as well as requirements on a spatial separation $r$ usually expressed in terms of its significance $r/\sigma_r$. This significance is calculated from the position errors of the two vertices which are determined from the error matrices of the tracks used. Fig. 3 shows the invariant mass spectra obtained for 3 different data samples using a significance cut of about $3\sigma$. These comprise the decay channels $\Omega^- \pi^+ \pi^- \pi^+$ (henceforth referred to as sample A) and $\Xi^- K^- \pi^+ \pi^+$. The latter channel has been split into two independent samples, one using only $K^-$ well identified in the RICH (sample C), the other one using the complementary sample (with $K^-$ below the Cherenkov threshold or outside the RICH acceptance) coming from interactions in the carbon target (sample B). The reason for the exclusion of interactions in the other targets from the last sample will be explained below. Although $\Omega_c^0$ decays have been reconstructed in more decay channels [24, 25] here we only present those final states used for the lifetime measurement, selected by statistical significance and life time resolution. Table 1 gives a summary of the signals from Fig. 3.

| channel | mass [MeV/c²] | width [MeV/c²] | S/B |
|---|---|---|---|
| A) $\Omega^- \pi^+ \pi^- \pi^+$ | 2707.5±3.9 | 8.3±2.9 | 23/18 |
| B) $\Xi^- K^- \pi^+ \pi^+$ (Carbon) | 2720.0±3.4 | 8.8±3.4 | 62/162 |
| C) $\Xi^- K^- \pi^+ \pi^+$ (high K-momenta) | 2719.8±4.7 | 8.9±3.0 | 27/50 |

Table 1: Properties of the $\Omega_c^0$ signals from Fig. 3 obtained from a likelihood fit to the data. The errors quoted are statistical only. Systematic errors in particular on the mass determinations are currently under study. The observed differences in mass for the three measurements have, however, no influence on the determination of the $\Omega_c^0$ lifetime. S/B is given for a mass windows of ±15 MeV/c².

Topological selections such as the $3\sigma$ cut mentioned above usually cause distortions of the decay distributions which are not easy to represent by Monte Carlo simulations. This is particularly true for very short lived particles. It therefore is the aim of this analysis to determine the lifetime of the $\Omega_c^0$ without such cuts.

In the following we present the evidence for a short but measurable lifetime of the $\Omega_c^0$ using all three data samples. Fig. 4a shows the distribution of the measured decay times

$$t = (x_{sec} - x_{prim})/\beta\gamma c$$

for events off the signal region (see Fig. 3) from sample A. No minimal separation between main and secondary vertex has been required. The distribution reflects the measurement accuracy obtained and can be fitted empirically using two Gaussians with widths of 31 and 69.5 fs with relative weights of 1:2. They are well reproduced in the simulation. In Fig. 4b we show the same distributions for events from the signal band which shows a shift towards positive values indicating a short but measurable lifetime for the $\Omega_c^0$. Fig. 5a shows the same distribution for sample B. Again, the distribution can be described by two Gaussians of width 48 and 125 fs and relative weights of about 4:1. Fig. 5b gives the mass distribution for these combinations using a separation cut of $3\sigma$ for positive and negative vertex separations. We clearly see that a signal only exists for positive (physical) lifetimes and that the background is similiar in both cases at the level of 5-10%. This again is clear evidence for a finite lifetime. Finally we show in Fig. 6 the invariant mass distributions for the sample C which contains $\Xi^- K^- \pi^+ \pi^+$ combinations with a kaon positively identified in the RICH. The distributions are plotted for different minimal decay lengths required of 0, 50, 100 and 150 fs. The short but measurable lifetime of the $\Omega_c^0$ is clearly seen. For this data sample the resolution on the decay times is again described by two Gaussians ($\sigma_1$=45 fs, $\sigma_2$=105 fs) with relative weights of 4:1.

The origin of the different values for the time resolution of different data samples lies in the different underlying momentum spectra of the $\Omega_c^0$ candidates. Sample B exhibiting the longest tails shows the lowest average momentum.

Samples with low energy kaons from interactions in copper have not been considered. Due to the large contribution of a component with $\sigma$ =150 fs no clear signal could be obtained without additional, topological selection criteria such as vertex isolation. This wide distribution can be explained by the large amount of multiple scattering to which the low momentum secondaries are subjected since the copper target is the most upstream one.

# 4   Determination of the Lifetime

For the determination of the lifetime, a sample is chosen from a mass window around the peak centre ('signal band'). A second sample from mass windows away from the peak ([2500,2650] MeV/c$^2$ and [2750,2900] MeV/c$^2$) is used to study the background ('side band'). The mass window chosen for the signal band is $\pm 15$ MeV/c$^2$ corresponding to about $\pm 2\sigma$ obtained from the fits to the mass spectra. The lifetime is determined from a maximum likelihood fit to the decay time ($t$) distribution of the signal band using a continuous likelihood method.

The likelihood function for each event consists of an exponential with lifetime $\tau$ convoluted with a resolution function $f_s(t)$ and a function $f_b(t)$ which describes the background:

$$\mathcal{L}_\tau(t) \;=\; \eta \frac{1}{N_\tau} \int_0^\infty f_s(t-\theta)\, e^{-\theta/\tau} d\theta \;+\; (1-\eta) f_b(t)$$

with the normalisation factor

$$N_\tau \;=\; \int_{t_{min}}^\infty \int_0^\infty f_s(t-\theta)\, e^{-\theta/\tau} d\theta dt$$

depending on the cut on the minimum accepted decay time. In this analysis we assumed that the time distribution of the background events reflects the time resolution and therefore we set $f_s(t)$=$f_b(t)$.

The resolution functions

$$f_{s,b}(t) \;=\; \frac{\alpha}{\sigma_1\sqrt{2\pi}} e^{-t^2/(2\sigma_1^2)} \;+\; \frac{1-\alpha}{\sigma_2\sqrt{2\pi}} e^{-t^2/(2\sigma_2^2)}$$

consist of two Gaussians for which the parameters are obtained from a fit to the side band distribution (see above). The background events in the signal band are assumed to have the same distribution as in the side bands.

The parameter $\eta$, giving the fraction of the signal in the mass window considered, is estimated from a fit to the observed mass distributions. Its value is varied in the lifetime fit under the constraint of a Poisson distribution which is determined by the number of events in signal and side bands.

The fit results are shown in Fig. 7. They are displayed as a function of the lowest lifetime $t_{min}$ still used in the fit. The errors are the statistical errors of the fits. The results for the different samples agree very well over the full range of $t_{min}$. The fit results are also shown in Table 2. The average values for the lifetime are obtained from a simultaneous likelihood fit to all three samples.

Table 2: Lifetimes measured in different samples as a function of the lowest decay time $t_{min}$ used in the fit.

| sample | $t_{min}$ [fs] | $\tau$ [fs] | $\eta$ | Combinations in signal band |
|---|---|---|---|---|
| $\Omega^- \pi^+ \pi^- \pi^+$ | all | $32\,^{+15}_{-12}$ | $.27 \pm .08$ | 112 |
| | 0 | $23\,^{+15}_{-14}$ | $.41 \pm .10$ | 68 |
| | 25 | $26\,^{+16}_{-13}$ | $.49 \pm .11$ | 43 |
| | 50 | $29\,^{+25}_{-22}$ | $.45 \pm .14$ | 26 |
| | 75 | $50\,^{+56}_{-26}$ | $.36 \pm .18$ | 11 |
| | 100 | $63\,^{+90}_{-34}$ | $.39 \pm .24$ | 4 |
| $\Xi^- K^- \pi^+ \pi^+$ from Carbon | all | $29\,^{+17}_{-17}$ | fixed at .08 | 1158 |
| | 0 | $20\,^{+13}_{-15}$ | $.18 \pm .04$ | 594 |
| | 25 | $36\,^{+15}_{-15}$ | $.21 \pm .04$ | 374 |
| | 50 | $57\,^{+15}_{-13}$ | $.25 \pm .05$ | 207 |
| | 75 | $63\,^{+17}_{-14}$ | $.30 \pm .06$ | 128 |
| | 100 | $74\,^{+21}_{-16}$ | $.34 \pm .07$ | 64 |
| $\Xi^- K^- \pi^+ \pi^+$ K ident. in RICH | all | $58\,^{+27}_{-22}$ | $.14 \pm .05$ | 259 |
| | 0 | $38\,^{+30}_{-28}$ | $.25 \pm .07$ | 151 |
| | 25 | $59\,^{+26}_{-20}$ | $.29 \pm .08$ | 93 |
| | 50 | $62\,^{+26}_{-19}$ | $.40 \pm .09$ | 56 |
| | 75 | $70\,^{+34}_{-22}$ | $.45 \pm .11$ | 34 |
| | 100 | $83\,^{+50}_{-28}$ | $.48 \pm .14$ | 20 |
| all channels | all | $28\,^{+9}_{-9}$ | | |
| | 0 | $25\,^{+9}_{-9}$ | | |
| | 25 | $37\,^{+12}_{-11}$ | | |
| | 50 | $55\,^{+13}_{-10}$ | | |
| | 75 | $64\,^{+15}_{-12}$ | | |
| | 100 | $77\,^{+19}_{-15}$ | | |

## Studies of systematic errors

In the following we discuss various sources of systematic errors. The strongest systematic influence on the value of the lifetime comes from the variation of the lower decay time cut (see Fig. 7). We have investigated this effect with simulated events passing the same selection criteria. A similar systematic behaviour was observed, however of smaller amplitude. We also observe a systematic downward shift of the reconstructed main vertex which, however, shows little influence on $\tau$ for values of $t_{min} \geq 50$ fs. These shifts can be compensated by using Gaussians not centred at the origin. However, we do not correct our data for this effect but attribute a systematic error of $\Delta\tau = 28$ fs, allowing for the lowest and highest lifetime value observed to be included at their $1\sigma$ level.

Another source of uncertainty comes from the parametrisation of the resolution function. For each decay it is chosen to be the sum of two Gaussians fitted to the time distributions in the sidebands (see above). Since this function serves two purposes, as indicated by distinguishing between $f_s(t)$ and $f_b(t)$, we have investigated them separately. If the wider Gaussian is omitted completely in the resolution function $f_s(t)$ for the signal events, but retained for $f_b(t)$, the lifetime increases by 7-9 fs for $t_{min}$ in the range of 0 to 100 fs. The Gaussians do not fully describe the background time distribution at t > 200 fs for the $\Omega^- \pi^+ \pi^- \pi^+$ mode and at t > 300 fs for the $\Xi^- K^- \pi^+ \pi^+$ mode. A bias from this should show up in a variation of the lifetime if an upper limit $t_{max}$ is applied to the fitted time spectra. For our chosen value $t_{min} = 50$ fs, the lifetime varies by $\pm 10$ fs if $t_{max}$ is varied from $+\infty$ down to below 200 fs. We therefore attribute a total systematic uncertainty of $^{+13}_{-10}$ fs to account for possible errors connected with the resolution function or large timings. Secondly we have investigated the influence of the description of the background by $f_b(t)$. Variations in the fraction of the long living component by $\pm 10$-$20\%$ cause variations in the fitted lifetime, in particular for small values of $t_{min}$. At large values ($t_{min} \geq 50$ fs) this effect becomes small ($\pm 5$ fs). The sample most affected is sample B which exhibits the largest tails in the resolution function. The variations are of the order of the statistical error for $t_{min} \leq 25$ fs. However, the influence on the averaged lifetime is small ($\pm 4$ fs). The sample $\Omega^- \pi^+ \pi^- \pi^+$, on the other hand, shows only little dependence ($\pm 5$ fs for variations of about 20% on the fraction of the long living component), as the resolution function exhibits shorter tails. The same holds for sample C where only the data point with $t_{min}=0$ shows some significant variation. The total systematic error attributed to uncertainties in $f_b(t)$ is $\pm 5$ fs.

Since the resolution for secondary vertices depends on the momentum of the daughter particles we have checked possible biases stemming from the momentum spectrum. We have subdivided our $\Omega_c^0$ samples rejecting about 30% of the events from low and high energy candidates, respectively, and repeated our analysis. We have found a small systematic dependence on the momentum of the order of $\pm 5$ fs.

In addition, we have investigated the sensitivity of the fits to the choice of the mass window chosen for the signal. Only a small variation ($\pm 5$ fs) is observed varying the mass window between $\pm 10$ and $\pm 20$ MeV/c$^2$.

From the present understanding of our detector, from our reconstruction algorithms and analysis cuts we do not observe a significant influence on the lifetime: biases of the resulting reconstruction efficiencies are small thus we omit the application of a correction function to our measured decay time distributions. Finally we investigated a possible dependence on the choice of the background windows chosen as well as on the $\chi^2$ cuts used for the secondary vertices. No significant influence was observed. In particular the timing distributions for the background windows below and above the $\Omega_c^0$ mass are alike.

Since the influence of systematic effects becomes smaller for larger lifetime cuts ($t_{min} \geq 50$ fs) we choose the data point at $t_{min}=50$ fs as central value since it has the best statistical significance. This choice leads to an asymmetric systematic error evaluation due to the dependence of $\tau$ on $t_{min}$. The total systematic error is estimated to be $^{+18}_{-23}$ fs.

## 5    Conclusion

We have measured the lifetime of the $\Omega_c^0$ obtaining the final result:

$$\tau(\Omega_c^0) = 55^{+13}_{-11}(stat.)^{+18}_{-23}(syst.) \text{ fs.}$$

This value makes the $\Omega_c^0$ the shortest living weakly decaying particle observed so far. The short lifetime has to be attributed to the properties of the final state containing three strange quarks and shows that the values of charmed hadron lifetimes depend largely on the structure of the baryon.

## 6    Acknowledgement

It is a pleasure to thank J. Zimmer and G. Konorova for their unstinted support in construction and setting up of the experiment. The experiment would not have been possible without the support from the technical staff of our institutes. We are also indebted to the staff of the Omega spectrometer group for their help and support during this experiment. We would also like to thank the CERN EBS group for their work on the hyperon beam line and the CERN accelerator group for their continuous efforts to provide good and stable beam conditions. Discussions with B. Stech on the strong effects in charm decays are gratefully acknowledged.

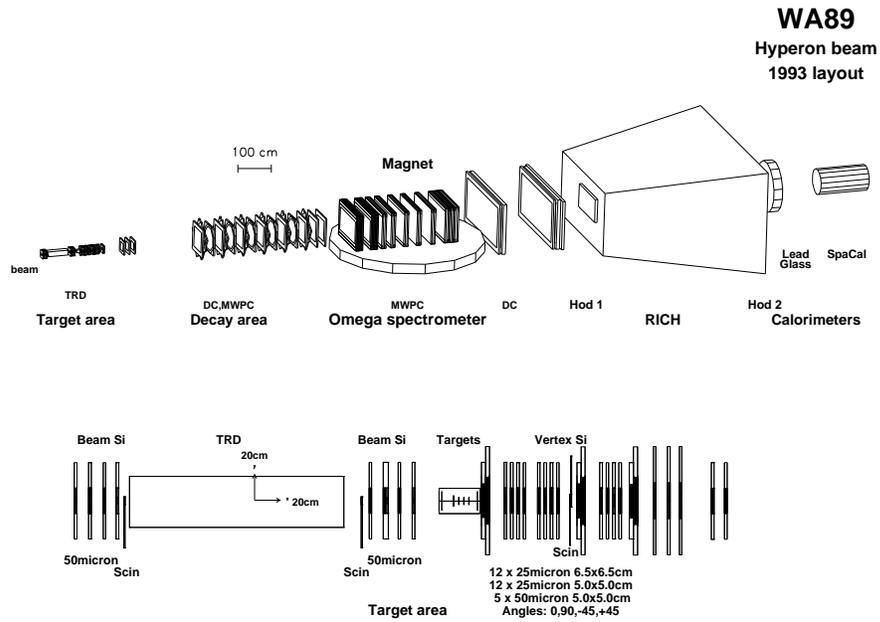

Figure 1: WA89 experimental set-up in 1993

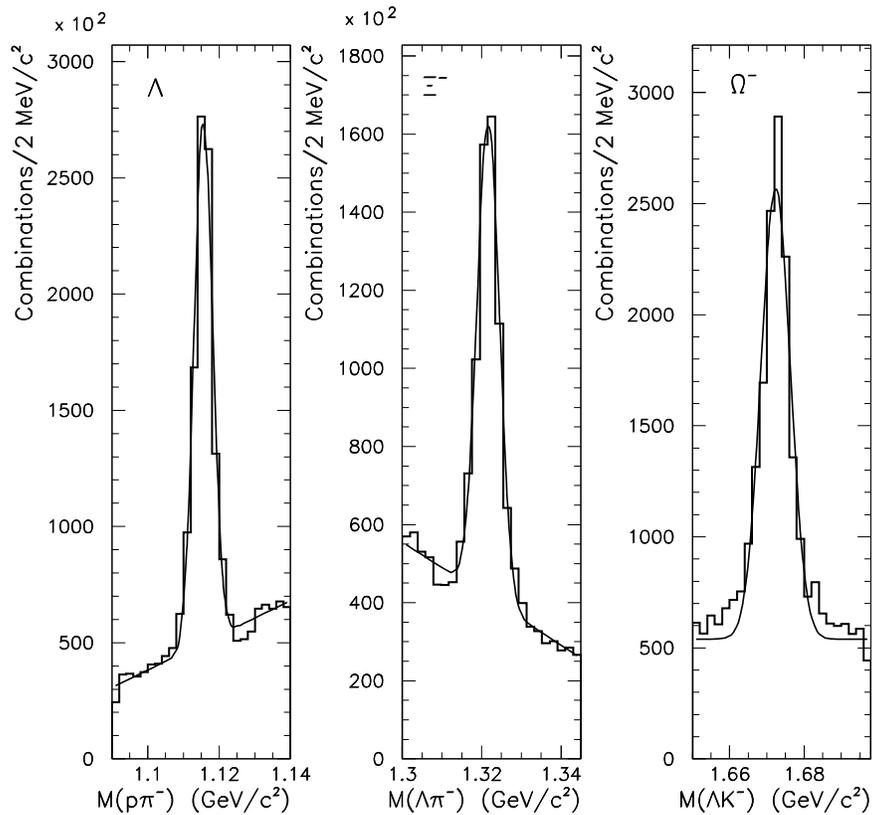

Figure 2: Examples for mass distributions of reconstructed strange particles.

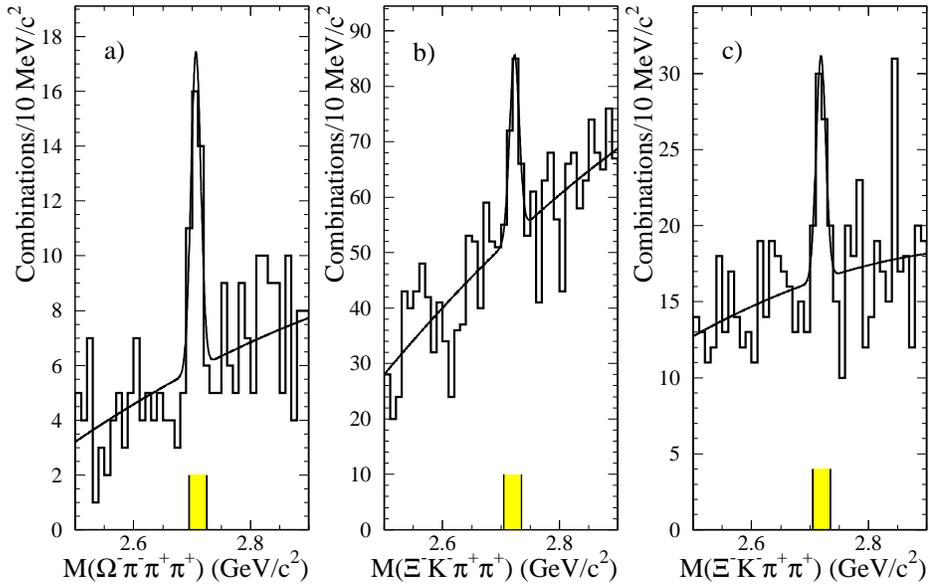

Figure 3: Mass distributions for three different final states: a) $\Omega^-\pi^+\pi^-\pi^+$, b) $\Xi^- K^- \pi^+\pi^+$ from carbon, c) $\Xi^- K^- \pi^+\pi^+$ from all targets with positively RICH identified kaon. The shaded region denotes the signal band.

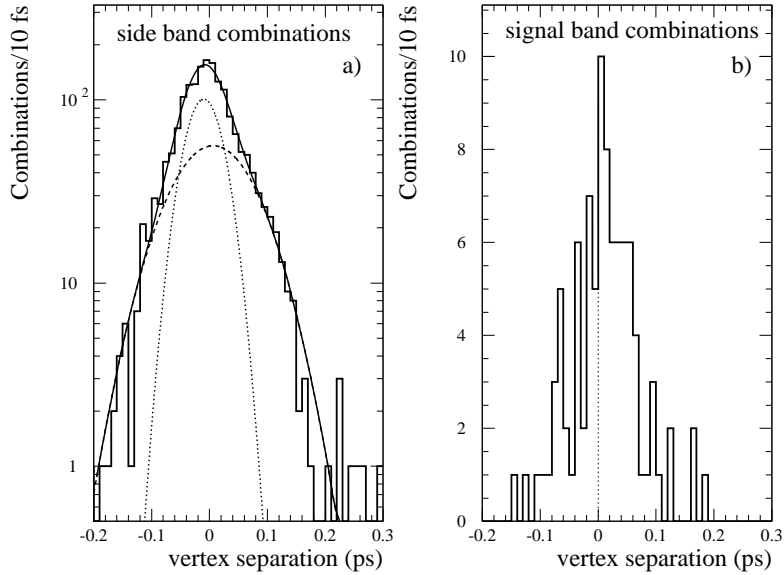

Figure 4: $\Omega^-\pi^+\pi^-\pi^+$ sample : vertex separation (ps) for a) background events, b) for signal band.

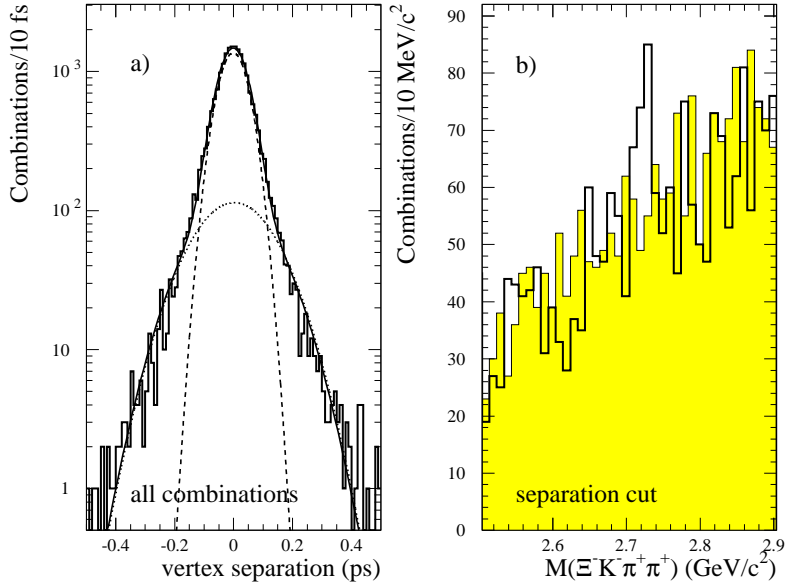

Figure 5: $\Xi^- K^- \pi^+ \pi^+$ sample in carbon: a) vertex separation for background events (ps) b) $\Xi^- K^- \pi^+ \pi^+$ mass spectrum after requiring $3\sigma$ separation of the two vertices (negative separations: filled histogram, positive separations: open histogram).

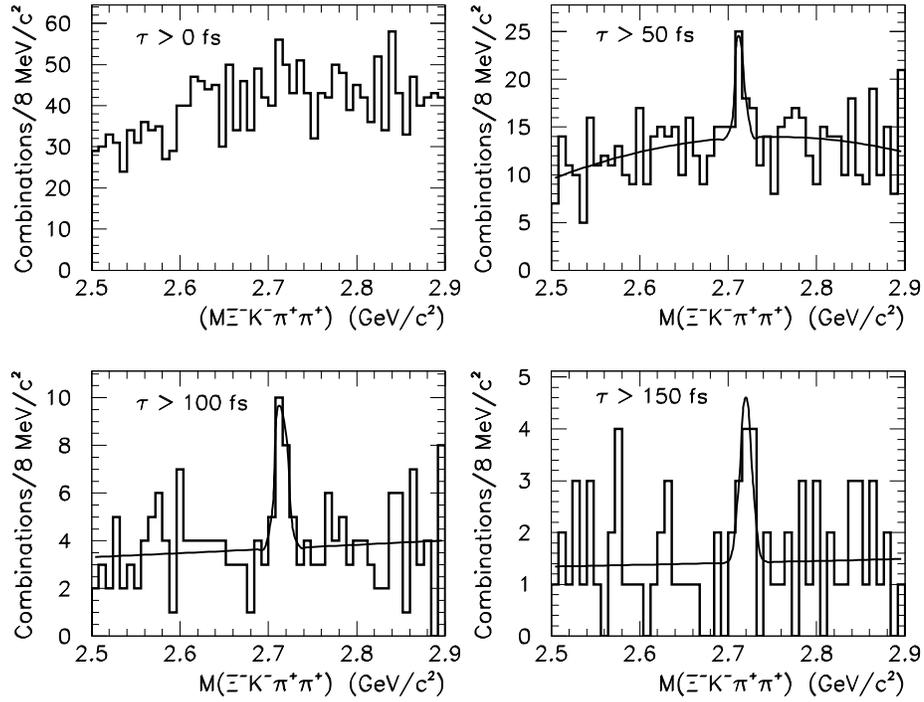

Figure 6: invariant mass distributions for the $\Xi^- K^- \pi^+ \pi^+$ sample with positively identified kaons using different lifetime cuts for the charm candidates.

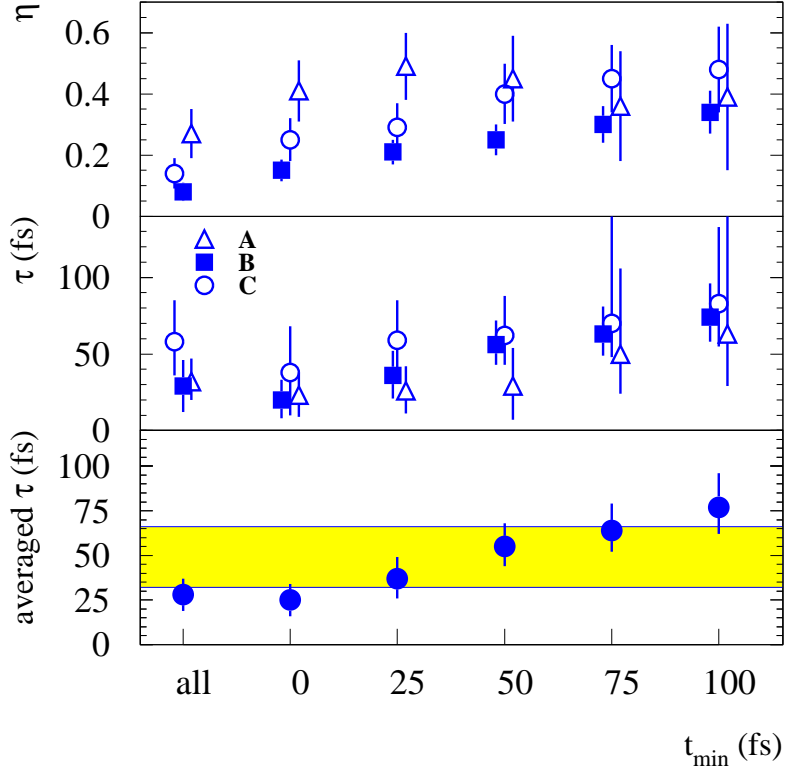

Figure 7: Results of the likelihood fits (see text) for the different samples and their dependence on the lower decay time cut $t_{min}$ used in the fit. The shaded area represents the size of the full systematic error. Sample A) denotes $\Omega^- \pi^+ \pi^- \pi^+$, sample B) $\Xi^- K^- \pi^+ \pi^+$ from the carbon target and sample C) $\Xi^- K^- \pi^+ \pi^+$ with identified $K^-$.